\documentclass[%
 reprint,
 amsmath,amssymb,
 aip,
 apl,
]{revtex4-1}

\usepackage{graphicx}
\usepackage{amsmath}
\usepackage{amssymb}
\usepackage{color}
\usepackage{verbatim}
\usepackage{url}
\usepackage{bm}
\usepackage{multirow}
\usepackage{booktabs}
\usepackage{rotating}

\begin{document}

\title{Analytical and numerical modeling of capacitance voltage characteristics of organic solar cells  }% Force line breaks with \\
\author{Prashanth Kumar Manda}
\email{ee12s013@ee.iitm.ac.in}
 
\author{Soumya Dutta}%
\email{s.dutta@ee.iitm.ac.in}
\affiliation{ 
Department of Electrical Engineering, Indian Institute of Technology Madras,Chennai,600036,India %\\This line break forced with \textbackslash\textbackslash
}%

\date{\today}% It is always \today, today,

\begin{abstract}

We report a charge based model to establish the analytical equation, describing the nature of capacitance-voltage ($C$-$V$) characteristics of organic solar cells under dark condition (or organic diodes) over a wide range of voltage, extending from deep reverse bias to well above the built-in potential. The present model enables to explain the unusual peak in $C$-$V$ characteristics of these devices, as observed experimentally by various groups. The results are further validated using self-consistent numerical simulation. Finally, we discuss the discrepancy and limitation of using Mott-Schottky relation to extract the built-in potential, doping density etc. in organic solar cells/diodes.

\end{abstract}

\maketitle

Significant improvement in efficiency of organic solar cell, especially over the last five years has shown a promising direction in third generation photovoltaic research as far as performance per cost ratio is concerned. Considerable research efforts are being carried out both experimentally and theoretically to improve the device performance and to understand the device physics. The analysis of experimentally observed current density-voltage ($J$-$V$) characteristics and the capacitance-voltage ($C$-$V$) characteristics of organic diodes, using numerical simulations and analytical equations have been reported by several research groups. Further, several attempts have been made to extract the crucial device parameters such as unintentional doping density ($N_A$), built-in potential ($V_{bi}$) and depletion width, upon applying Mott-Schottky (MS) relationship to the $C$-$V$ characteristics, as adopted in traditional silicon diodes.\cite{Sze,Capofpentacene} However, the understanding of device operation in the same context of traditional silicon diodes is still conflicting and debatable too.

The validity of application of MS relationship in organic diodes to extract parameters is questionable due to the realization of several discrepancies that have been reported recently. First, the extracted doping concentration (using MS relationship) of the diodes, fabricated in atmospheric condition, has been observed to be apparently the same ($N_A\approx 10^{16}$ cm$^{-3}$) as that of the diodes, fabricated under inert ambient (inside glove box). The sources of such a high doping density even after fabricating the device inside a glove box is still not clear. Secondly, both $V_{bi}$ and $N_A$, extracted using MS relationship, have been observed to vary with the thickness of the semiconductor,\cite{SensitivityofMS,Vbi&validityofMS} which is not justified anywhere. Recent simulation studies have shown that the parameters, extracted using MS relationship for organic devices, are erroneous, which is the consequence of analytical artifact of applying MS relationship.\cite{Nigam1,Nigam2roleofinject,SensitivityofMS}

The $C$-$V$ characteristics of organic diodes appear different from those of conventional Schottky or \textit{p}-\textit{n} junction diodes, especially in the large forward bias region.  Under reverse bias, the capacitance remains almost constant, with a value close to the geometrical capacitance. Upon applying forward bias, the capacitance gradually increases up to a maximum value, followed by a decrease with the increase in voltage. This leads to a peak and pseudo symmetrical $C$-$V$ characteristics unlike conventional diodes. The pseudo symmetric nature of $C$-$V$ characteristic of organic diode around the peak has been speculated in different ways elsewhere.\cite{Vbi&validityofMS,DiffusivecapIITK,Bisquert200857} One proposition is based on the charging of trap states (before the peak) and neutralization of traps (after the peak).\cite{CVwithTraps} Another proposition is based on accumulation of the injected majority carriers (before the peak), followed by recombination of injected minority carriers (after the peak).\cite{CVPLED.Vishal} Nevertheless, there are no experimental proofs to support the propositions. Recently an analytical equation, based on space charge has been derived to explain the capacitance behavior before the onset of the capacitance peak.\cite{DiffusivecapIITK} However, there exists no single analytical equation, which can elucidate the nature of $C$-$V$ characteristics of organic diodes, including the peak. Towards this end, the understanding of device physics, based on coherent physical model, is indispensable to standardize the device parameters, thereby improving the device performance. 

In this letter, we derive the analytical equation for capacitance (consider capacitance density hereafter) of organic solar cell under dark condition using the charge density therein. The analytical results are validated using Metal-Insulator-Metal (MIM) model based numerical device simulation and are further verified by comparing with the experimental results, reported earlier.\cite{Vbi&validityofMS} The derived conduction capacitance ($C_{cond}$), using the present model, can explain the origin of the peak in $C$-$V$ characteristics and its pseudo symmetrical nature. The total capacitance is modeled as a parallel combination of geometrical capacitance and conduction capacitance that can explain the analytical artifact of applying the MS relation. 
%%%%%%%%%%%%%%%%%%%%%%%%%%%%%%%%%%%%%%%%%%%%%%%%%%%%%%%%%%%%%%%%%%%%%%%%%%%%%%%%%%%%%%
%%%%%%%%%%%%%%%%%%%%%%%%%%%%%%%%%% Figure-1%%%%%%%%%%%%%%%%%%%%%%%%%%%%%%%%%%%%%%%%%%%
\begin{figure}[t]
\centering
\includegraphics[scale=1]{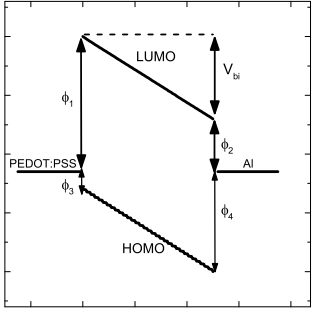}
\caption{Band diagram of PEDOT:PSS/P3HT:PCBM/Al under equilibrium.}\label{<C2Vth>}
\end{figure}
%%%%%%%%%%%%%%%%%%%%%%%%%%%%%%%%%%%%%%%%%%%%%%%%%%%%%%%%%%%%%%%%%%%%%%%%%%%%%%%%%%%%%%

To start with, we consider a standard organic solar cell, based on poly(3-hexylthiophene) (P3HT): phenyl-C61-butyric acid methyl ester (PCBM), as active medium, aluminum (Al) as cathode and indium tin oxide (ITO)/ poly(3,4-ethylenedioxythiophene):polystyrene sulfonate (PEDOT:PSS) as anode, as illustrated in Fig.\ref{<C2Vth>}(a). We adopt MIM model for device simulation that has been widely used by several groups to realize the experimental characteristics.\cite{SensitivityofMS,Coupledoptoelectronic,Koster} This model considers the organic blend material as an effective medium of insulator. The charge transport, continuity and Poisson's equations are discretized by Scharfetter-Gummel discretization scheme, and are solved self consistently by using Gummel's method.\cite{Scharfetter} In particular, contact metals (cathode and anode) inject the carriers (by thermionic emission process) into the device and creates space charge. The injected space charge is sufficiently low to perturb the band diagram. Hence the band diagram remains similar to that of an insulator kept between two different metals, as shown in Fig.\ref{<C2Vth>}. As a consequence, there exists an equilibrium that leads to the formation of built-in electric field and the carrier distribution inside the device. The injected charge carriers, even though are less in amount, play a major role in the charge dynamics of the device.\cite{Simmons} The strength of the electric field and the magnitude of the injected charge carrier density depend on the work-function of the contact metals and the thickness of the active material. The equilibrium band bending for P3HT:PCBM solar cell is shown in Fig.\ref{<C2Vth>}b where, the built-in potential, as defined in this study, $V_{bi}= (\phi_1-\phi_2)/q= (\phi_4-\phi_3)/q$ is the difference in the work-function of the metals (cathode and anode) and $\phi_1$($\phi_3$), $\phi_2$($\phi_4$) are electron (hole) injection barriers at anode and cathode respectively.   

The analytical equations are established, based on the following assumptions: (i) The blend material is trap free with constant carrier mobilities ($\mu_n,\mu_p$) with respect to the applied voltage $V$, (ii) the electric field within the active material of thickness $d$ is uniform ($-(V_{bi}-V)/d$), (iii) there is no recombination. The carrier density profiles are derived by solving the Poisson's equation, coupled with the charge transport and the continuity equations and using the aforementioned boundary conditions. The derived spatial distribution of electron concentration is given as 
%%%%%%%%%%%%%%%%%%%%%%%%%%%%%%%%%%%%%%%%%%%%%%%%%%%%%%%%%%%%%%%%%%%%%%%%%%%%%%%%%%%%%%
%%%%%%%%%%%%%%%%%%%%%%%%%%%%%%%%%% Equation-1 %%%%%%%%%%%%%%%%%%%%%%%%%%%%%%%%%%%%%%%%%%%
\begin{equation}
\begin{array}{l} 
n(x,V_a)=\frac{n_d-n_0\exp\left(\frac{V_{bi}-V}{V_t}\right)+\left(n_0-n_d\right)\exp\left(\frac{V_{bi}-V}{V_t}\frac{x}{d}\right)}{1-\exp\left(\frac{V_{bi}-V}{V_t}\right)}\label{nden}
\end{array}
\end{equation}
where
\begin{equation*}
n_0=N_c\exp\left(-\frac{\phi_1}{qV_t}\right),n_d=N_c\exp\left(-\frac{\phi_2}{qV_t}\right)
\end{equation*}
  $V_t$ is the thermal voltage, $n_0$ and $n_d$ are electron concentrations at $x=0$ (anode) and $x=d$ (cathode) respectively. Similar expression for hole concentration $p(x$,$V$) can also be achieved in terms of $p_0$ and $p_d$, the hole concentrations at anode and cathode respectively.   

 The conduction capacitance ($C_{cond}$) can be calculated by differentiating the total charge density ($Q$) with respect to $V$, where the total charge density is calculated as
%%%%%%%%%%%%%%%%%%%%%%%%%%%%%%%%%%%%%%%%%%%%%%%%%%%%%%%%%%%%%%%%%%%%%%%%%%%%%%%%%%%%%%%%%%%%
\begin{align}
Q(V)=q\left[\int\limits_{0}^{d} n(x,V)\,dx+\int\limits_{0}^{d} p(x,V)\,dx\right]
\end{align}
%%%%%%%%%%%%%%%%%%%%%%%%%%%%%%%%%%%%%%%%%%%%%%%%%%%%%%%%%%%%%%%%%%%%%%%%%%%%%%%%%%%%%%%%%%%%
\begin{multline}
Q(V)=qd\left( p_d+n_0-\left(p_0-p_d-n_0+n_d\right)\left[\frac{V_t}{V_{bi}-V}\right]\right)\\
+qd\left(p_0-p_d-n_0+n_d\right)\left[\frac{1}{1-\exp\left(\frac{V_{bi}-V}{V_t}\right)}\right]
\label{Tcharge}
\end{multline}
%%%%%%%%%%%%%%%%%%%%%%%%%%%%%%%%%%%%%%%%%%%%%%%%%%%%%%%%%%%%%%%%%%%%%%%%%%%%%%%%%%%%%%%%%%%%
%%%%%%%%%%%%%%%%%%%%%%%%%%%% Figure-2 %%%%%%%%%%%%%%%%%%%%%%%%%%%%%%%%%%%%%%%%%%%
\begin{figure}[t]
\centering
\includegraphics[trim = 40mm 10mm 40mm 20mm,scale=0.25]{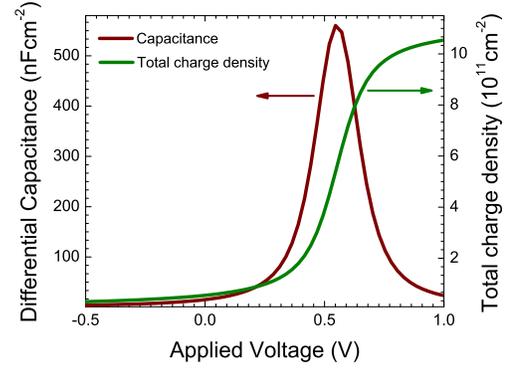}
\caption{Total charge density and conduction capacitance of 50 nm thick P3HT:PCBM solar cell.}\label{Tcharge&C}
\end{figure}
%%%%%%%%%%%%%%%%%%%%%%%%%%%%%%%%%%%%%%%%%%%%%%%%%%%%%%%%%%%%%%%%%%%%%%%%%%%%%%%%%%%%%%%%%%%%%%
%%%%%%%%%%%%%%%%%%%%%%%%%%%%%%%%%%%%%%%%%%%%%%%%%%%%%%%%%%%%%%%%%%%%%%%%%%%%%%%%%%%%%%%%%%%%
Hence the conduction capacitance ($C_{cond}$)
\begin{align}
C_{cond}=\frac{dQ}{dV} 
\end{align}
%%%%%%%%%%%%%%%%%%%%%%%%%%%%%%%%%%%%%%%%%%%%%%%%%%%%%%%%%%%%%%%%%%%%%%%%%%%%%%%%%%%%%%%%%%%%
\begin{multline}
C_{cond}= qd\frac{p_0-p_d-n_0+n_d}{V_t}\left(\frac{V_t}{V_{bi}-V}\right)^2\\
-qd\frac{p_0-p_d-n_0+n_d}{V_t}\frac{\exp\left(\frac{V_{bi}-V}{V_t}\right)}{\left(1-\exp\left(\frac{V_{bi}-V}{V_t}\right)\right)^2}\label{Ccond}
\end{multline}

The variation of total charge density and conduction capacitance with applied voltage is depicted in Fig.\ref{Tcharge&C}, following Eq.\ref{Tcharge} and Eq.\ref{Ccond}. The total charge density gradually increases with applied voltage, spanning from reverse bias to low forward bias and then increases rapidly around $V_{bi}$, followed by a nearly saturation at higher voltages. This leads to a peak around $V_{bi}$ and pseudo symmetric nature of $C$-$V$ characteristics, as shown in Fig.2.

The voltage, applied across an organic semiconductor, drops throughout the layer, similar to the case of parallel plate capacitor or a fully depleted semiconductor kept in between two metals, as reported earlier.\cite{Capofpentacene} This results in the appearance of geometrical capacitance, which does not depend on the polarity and magnitude of applied voltage or metal work functions but solely depends on the thickness of the active material. Hence the total capacitance($C$) is modeled as a parallel combination of conduction capacitance and geometrical capacitance and is expressed as  

%%%%%%%%%%%%%%%%%%%%%%%%%%%%%%%%%%%%%%%%%%%%%%%%%%%%%%%%%%%%%%%%%%%%%%%%%%%%%%%%%%%%%%%%%%%%%%%%
\begin{equation}
C=\frac{\epsilon_0\epsilon_r}{d}+C_{cond}\label{TotalC}
\end{equation}
%%%%%%%%%%%%%%%%%%%%%%%%%%%%%%%%%%%%%%%%%%%%%%%%%%%%%%%%%%%%%%%%%%%%%%%%%%%%%%%%%%%%%%%%%%%%%%%%
where $\epsilon_0$, $\epsilon_r$ are permittivity of free space and dielectric constant of the active semiconductor respectively.     

%%%%%%%%%%%%%%%%%%%%%%%%%%%%%%%%%%%%%%%%%%%%%%%%%%%%%%%%%%%%%%%%%%%%%%%%%%%%%%%%%%%%%%%%%%%%%%%%%
%%%%%%%%%%%%%%%%%%%%%%%%%%%%%%%%% Figure-3 %%%%%%%%%%%%%%%%%%%%%%%%%%%%%%%%%%%%%%%%%%%%%%%%%%%%%
\begin{figure}[t]
\centering
\includegraphics[trim = 40mm 10mm 40mm 20mm,scale=0.25]{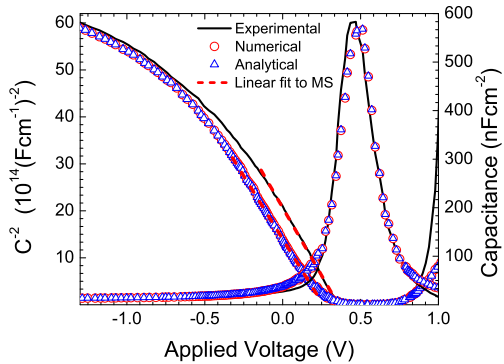}
\caption{Experimental,\citep{Vbi&validityofMS} numerical and analytical $C^{-2}$-$V$ and $C$-$V$ characteristics of P3HT:PCBM solar cell with $E_g$=1.1 eV, $\phi_1$=1.0684 eV, $\phi_3$=0.0316 eV, $V_{bi}$=0.5388 V, $\mu_n=\mu_p=10^{-4}$ cm$^2$V$^{-1}$s$^{-1}$, $N_C=1\times10^{17}$ cm$^{-3}$, $N_V=1.5\times10^{17}$ cm$^{-3}$, $\epsilon_r$=3.3, $d$=250 nm, $V_t$=26 mV}\label{CV-C2V}
\end{figure}
%%%%%%%%%%%%%%%%%%%%%%%%%%%%%%%%%%%%%%%%%%%%%%%%%%%%%%%%%%%%%%%%%%%%%%%%%%%%%%%%%%%%%%%%%%%%%%%%%

The total analytical capacitance, obtained from Eq.\ref{TotalC}, shows a good match with our numerical simulation results and the experimental $C$-$V$ characteristics, reported by Dyakonov et.al.,\cite{Vbi&validityofMS} as shown in Fig.\ref{CV-C2V}. The similar characteristics have also been reported by several groups. \cite{DiffusivecapIITK,Bisquert200857,DtrmOfInjBarr,CVwithvarCathode,SensitivityofMS,CV65} In deep reverse bias, conduction capacitance does not contribute,
resulting total capacitance to be equal to the geometrical capacitance. Upon applying small forward voltage, conduction capacitance dominates over the geometrical
capacitance, resulting the appearance of unusual peak in $C$-$V$ characteristics, as mentioned above. 

\begin{figure}[t]
\centering
\includegraphics[trim = 40mm 10mm 40mm 20mm,scale=0.25]{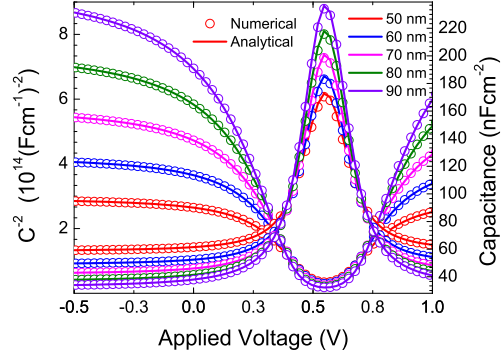}
\caption{$C^{-2}$-$V$ and $C$-$V$ characteristics of P3HT:PCBM solar cell with different thicknesses.}\label{DCV1}
\end{figure}

From Eq.(\ref{Ccond}), it is clear that the conduction capacitance depends on applied voltage, magnitude of injection barriers and the work-function difference. This supports the experimental results, where the position and magnitude of the capacitance peak has been observed to be dependent on the type of the metals, used for contacts.\cite{DtrmOfInjBarr,CVwithvarCathode,SensitivityofMS} Our numerical and analytical models are also able to explain the thickness dependency of capacitance (Fig.\ref{DCV1}), as observed experimentally. Typically, with increase in thickness, the magnitude of the minimum capacitance decreases due to the decrease in geometrical capacitance, whereas the maximum capacitance increases due to the linear increase of conduction capacitance with thickness. However, the position of capacitance peak is observed to remain same at the voltage equal to built-in potential. It corroborates that the built-in potential doesn't depend on the thickness, as observed in case of conventional diodes.\cite{Sze}
%%%%%%%%%%%%%%%%%%%%%%%%%%%%%%%%%%%%%%%%%%%%%%%%%%%%%%%%%%%%%%%%%%%%%%%%%%%%%%%%%%%%%%%%%%%%%%%%%
%%%%%%%%%%%%%%%%%%%%%%%%%%%%%%%% Figure -4 %%%%%%%%%%%%%%%%%%%%%%%%%%%%%%%%%%%%%%%%%%%%%%%%

%%%%%%%%%%%%%%%%%%%%%%%%%%%%%%%%%%%%%%%%%%%%%%%%%%%%%%%%%%%%%%%%%%%%%%%%%%%%%%%%%%%%%%%%%%%%%%%%%

 Recent reports on the numerical simulations of Metal-Insulator-Semiconductor (MIS) diode structure has shown that the capacitance is influenced by the injection barrier heights but not by the mobility of charge carriers.\cite{Nigam1,Nigam2roleofinject} Similar effects can be expected in organic diode too, as reflected in Eq.(\ref{Ccond}). It is to be noted that for small forward bias ($V<V_{bi}$), the non-exponential part dominates, showing linear relationship between $C^{-2}$ and $V$. This resembles MS relation in conventional \textit{p}-\textit{n} junction or Schottky diode though the proportionality factor, in case of organic diode, depends on the the thickness of the semiconductor and the charge densities at the contacts, unlike the other counterpart. 

In order to validate the relevance of MS relation in organic diode, we employ a straight line fitting to $C^{-2}$-$V$ characteristics, obtained from two different experimental results for active layers of thicknesses 65 nm \cite{CV65} and 250 nm .\cite{Vbi&validityofMS} The similar fitting is also applied to our numerical and analytical results for the corresponding thicknesses (Fig.\ref{CV-C2V}). The apparent doping density ($N_A$) and built-in potential ($V_{bi}$) are extracted from the slope and the voltage-intercept of fitted straight lines respectively, using MS relation 
%%%%%%%%%%%%%%%%%%%%%%%%%%%%%%%%%%%%%%%%%%%%%%%%%%%%%%%%%%%%%%%%%%%%%%%%%%%%%%%%%%%%%%%%%%%%%%%%%
\begin{equation}
\frac{1}{C^2}=\frac{2(V_{bi}-V)}{q\epsilon\epsilon_r N_A} \label{Mott-Schottky}
\end{equation}
%%%%%%%%%%%%%%%%%%%%%%%%%%%%%%%%%%%%%%%%%%%%%%%%%%%%%%%%%%%%%%%%%%%%%%%%%%%%%%%%%%%%%%%%%%%%%%%%%

The extracted parameters are tabulated in Table I, showing the thickness dependency of $N_A$ and $V_{bi}$. The later is also evident from Fig.\ref{DCV1}, as long as we follow MS relation. However, in the present context, we propose that the actual built-in potential of organic diode should be the applied voltage corresponding to the capacitance peak. This voltage is independent of thickness but depends only on the the type of the metal contacts, as shown in Fig.\ref{DCV1}. This is in agreement with the experimental results too.\cite{DtrmOfInjBarr,CVwithvarCathode}

\begin{table}[ht]
\centering
\begin{tabular}{|c|c|c|c|c|}
\hline 
\multirow{2}{*}{} & \multicolumn{2}{|c|}{65 nm} & \multicolumn{2}{|c|}{250 nm} \\ \cline{2-5}
 & $N_A$ (cm$^{-3}$) & $V_{bi}$ (V) & $N_A$ (cm$^{-3}$) & $V_{bi}$ (V) \\ 
\hline 
 Experimental & $5.44\times10^{16}$ & 0.59 & $6.96\times10^{15}$ & 0.35 \\ 
\hline 
Numerical & $3.92\times10^{16}$ & 0.60 & $1.08\times10^{16}$ & 0.30 \\ 
\hline 
Analytical & $3.93\times10^{16}$ & 0.61 & $1.07\times10^{16}$ & 0.30 \\ 
\hline 
\end{tabular} 
\caption{Comparison of extracted parameters, upon applying MS relation to experimental,\cite{Vbi&validityofMS,CV65}numerical and analytical results }
\label{table:1}
\end{table}

\begin{figure}[t]
\centering
\includegraphics[trim = 40mm 10mm 40mm 20mm,scale=0.25]{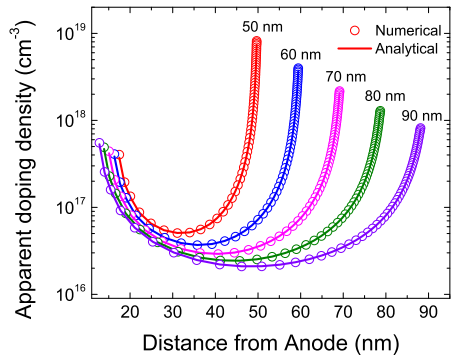}
\caption{Apparent doping density profile of organic diode with different active layer thicknesses}\label{DCV}
\end{figure}

To further investigate the thickness dependency of $N_A$, we plot the apparent doping density profile for different thickness of solar cells using the relation 
\begin{align}
N_A(x)=\frac{-2}{q\epsilon \epsilon_r}\left[\frac{\partial C^{-2}}{\partial V}\right]^{-1} \label{doping}
\end{align}  
where $x$ (the distance from anode) $=\epsilon \epsilon_r/C$ (Fig.\ref{DCV}). The apparent doping profiles, extracted from both analytical and the numerical simulation, are in good agreement, showing the consistency of these two approaches. Moreover, the nature of the profile is similar to the reported results, which has been attributed to unintentional doping. \cite{SensitivityofMS,Vbi&validityofMS} However, in our approach, the semiconductor is considered as purely intrinsic. Hence, the apparent doping profile is merely uncorrelated to the real doping.
%%%%%%%%%%%%%%%%%%%%%%%%%%%%%%%%%%%%%%%%%%%%%%%%%%%%%%%%%%%%%%%%%%%%%%%%%%%%%%%%%%%%%%%%%%%%%%%%%
%%%%%%%%%%%%%%%%%%%%%%%%%%%%%%%  Figure-5 %%%%%%%%%%%%%%%%%%%%%%%%%%%%%%%%%%%%%%%%%%%%
This is an unjustified artifact, arising due to forcible application of MS relation in organic diode without having any physical insight. 
%%%%%%%%%%%%%%%%%%%%%%%%%%%%%%%%%%%%%%%%%%%%%%%%%%%%%%%%%%%%%%%%%%%%%%%%%%%%%%%%%%%%%%%%%%%%%%%%%

In summary, we derive the analytical equations, based on coherent device physics, to interpret the typical $C$-$V$ characteristics of organic solar cell under dark condition (diode) comprehensively. The results are in good agreement with numerical simulation and the experimental results. Our studies elucidate the discrepancies like thickness dependency of built-in potential and apparent doping densities, arising due to inappropriate application of Mott-Schottky relation in organic diodes. Different nature of experimentally observed $C$-$V$ characteristics, compared to the traditional $p$-$n$ junction or Schottky diode, is attributed to the conduction capacitance, resulted from the injected space charge. Further extension of this approach towards finding the frequency dependent capacitance characteristics is our immediate interest of research, which enables to identify the role of traps and their extraction methodology in organic solar cells.
     
%\bibliography{References1}

\end{document}